\documentclass[12pt]{article}
\usepackage{amsmath}[1996/11/01]
\usepackage{fullpage}
\usepackage{amssymb,amsthm}

\newtheorem{thm}{Theorem}
\newtheorem{cor}[thm]{Corollary}
\newtheorem{lem}[thm]{Lemma}
\newtheorem{prop}[thm]{Proposition}
\theoremstyle{remark}

\newtheorem{rems}[thm]{Remark}
\newtheorem{trems}[thm]{Technical Remark}

\theoremstyle{definition}

\numberwithin{equation}{section}
\numberwithin{thm}{section}

\newcommand{\R}{\mathbb{R}}
\newcommand{\Z}{\mathcal{Z}}
\newcommand{\alphalm}{\alpha^{[\ell,m]}}
\newcommand{\alpham}{\alpha^{[0,m]}}
\newcommand{\pilm}{\pi^{[\ell,m]}}
\newcommand{\pim}{\pi^{[0,m]}}
\newcommand{\qm}{q^{[0,m]}}
\newcommand{\qlm}{q^{[\ell,m]}}
\newcommand{\piNK}{\pi_{N+K}}
\newcommand{\piNKone}{\pi_{N+K-1}}

\DeclareMathOperator{\Prob}{P}
\DeclareMathOperator{\sgn}{sgn}

\begin{document}

\title{{\bf Products and Ratios of Characteristic Polynomials of Random Hermitian Matrices}}
\author{{\bf Jinho Baik} \\
Department of Mathematics, Princeton University, \\ Princeton, New
Jersey, 08544, USA \\ jbaik@math.princeton.edu \\ Department of
Mathematics, University of Michigan \\ Ann Arbor, MI 48109, USA \\ \and {\bf Percy Deift} \\
Courant Institute of Mathematical Sciences, New York University \\
New York, NY 10012, USA \\ deift@cims.nyu.edu \\
School of Mathematics, Institute for Advanced Study \\
Princeton, NJ 08540, USA \and {\bf Eugene
Strahov} \\ Department of Matheamtical Sciences, Brunel University \\
Uxbridge, UB8 3PH, United Kingdom \\ Eugene.Strahov@brunel.ac.uk}

\date{\today}
\maketitle

\begin{abstract}
We present new and streamlined proofs of various formulae for
products and ratios of characteristic polynomials of random
Hermitian matrices that have appeared recently in the literature.
\end{abstract}

\newpage

\section{Introduction}\label{intro}

In random matrix theory, unitary ensembles of $N\times N$ matrices
$\{ H\}$ play a central role \cite{Mehta91}. Such ensembles are
described by a measure $d\alpha$ with finite moments $\int_\R
|x|^kd\alpha(x) <\infty$, $k=0,1,2,\cdots$, and the distribution
function for the eigenvalues $\{ x_i=x_i(H) \}$ of matrices $H$ in
the ensembles has the form
\begin{equation}\label{0.1}
  d\Prob_{\alpha,N}(x) = \frac1{Z_N} \Delta(x)^2 d\alpha(x)
\end{equation}
where $d\alpha(x) = \displaystyle \prod_{i=1}^N d\alpha(x_i)$,
$\Delta(x)=\displaystyle \prod_{N\ge i>j \ge 1}(x_i-x_j)$ is the
Vandermonde determinant for the $x_i$'s, and $Z_N=\int\cdots\int
\Delta(x)^2 d\alpha(x)$ is the normalization constant. The special
case $d\alpha(x)= e^{-x^2}dx$ is known as the Gaussian Unitary
Ensemble (GUE). For symmetric functions $f(x)=f(x_1,\cdots, x_N)$
of the $x_i$'s,
\begin{equation}\label{0.2}
  \bigl< f\bigr>_\alpha \equiv \frac1{Z_N} \int\cdots\int f(x)\Delta(x)^2
  d\alpha(x)
\end{equation}
denotes the average of $f$ with respect to $d\Prob_{\alpha,N}$.

Recently there has been considerable interest in the averages of
products and ratios of the characteristic polynomials
$D_N[\mu,H]=\displaystyle \prod_{i=1}^N(\mu-x_i(H))$ of random
matrices with respect to various ensembles. Such averages are
used, in particular, in making predictions about the moments of
the Riemann-zeta function, see \cite{KeatingS00, KeatingSH00,
KeatingHPO01} (circular ensembles) and \cite{BrezinH00} (unitary
ensembles). Many other uses are described, for example, in
\cite{AndreevS}, \cite{StrahovF1} and \cite{StrahovF2}.

By \eqref{0.2}, for unitary ensembles, such averages have the form
\begin{equation}\label{0.3}
  \biggl< \frac{\prod_{j=1}^K D_N[\mu_j,H]}{\prod_{j=1}^M D_N[\epsilon_j,H]} \biggr>_\alpha
  = \frac{1}{Z_N} \int\cdots\int
  \frac{\prod_{j=1}^K\prod_{i=1}^N(\mu_j-x_i)}{\prod_{j=1}^M\prod_{i=1}^N(\epsilon_j-x_i)}
  \Delta(x)^2 d\alpha(x).
\end{equation}
In this paper we consider certain explicit determinantal formulae
for \eqref{0.3} -- see \eqref{BrezinHikamiFormula},
\eqref{averageOM}, \eqref{AverageofRatios}, \eqref{D}, \eqref{E}
below. Formula \eqref{BrezinHikamiFormula} is due to Brezin and
Hikami \cite{BrezinH00} (see also \cite{MehtaN01}, and when all
the $\mu_j$'s are equal, see \cite{ForresterW}), whereas
\eqref{averageOM}, \eqref{AverageofRatios}, \eqref{D} and
\eqref{E} are due to Fyodorov and Strahov \cite{StrahovF1,
StrahovF2}. The papers \cite{StrahovF1, StrahovF2} also contain a
discussion of the history of these formulae. The formulae
\eqref{D} and \eqref{E} are particularly useful in proving
universality results for the ratios \eqref{0.3} in the Dyson limit
as $N\to\infty$ (see \cite{StrahovF2}). For a discussion of other
universality results, particularly the work of Brezin-Hikami and
Fyodorov in special cases, we again refer the reader to
\cite{StrahovF2}. The asymptotic analysis in \cite{StrahovF2} is
based on the reformulation of the orthogonal polynomial problem as
a Riemann-Hilbert problem by Fokas, Its and Kitaev \cite{FIK}. The
Riemann-Hilbert problem is then analyzed asymptotically using the
non-commutative steepest-descent method introduced by Deift and
Zhou \cite{DZ1}, and further developed with Venakides in
\cite{DVZ} to allow for fully non-linear oscillations, and in
\cite{DKMVZ2}, \cite{DKMVZ3}.

Our goal in this paper is to give new, streamlined proofs of
\eqref{BrezinHikamiFormula}-\eqref{E}, using only the properties
of orthogonal polynomials and a minimum of combinatorics. Along
the way we will also need an integral version of the classical
Binet-Cauchy formula due to C.Andr\'eief dating back to 1883 (see
Lemma \ref{LemmaChristoffel1} below).

Let $\pi_j(z)=x^j+\cdots$ denote the $j$th monic orthogonal
polynomial with respect to the measure $d\alpha$,
\begin{equation}\label{0.4}
  \int_{\R} \pi_j(x)\pi_k(x) d\alpha(x) = c_jc_k \delta_{jk},
  \qquad j,k\ge 0,
\end{equation}
where the norming constants $c_j$'s are positive. The key
observation in our approach is that for $K=1$ and $M=0$ in
\eqref{0.3}
\begin{equation}\label{0.5}
  \bigl< D_N[\mu,H] \bigr>_\alpha = \pi_N(\mu)
\end{equation}
(see \cite{Szego}). In our words, the orthogonal polynomial
$\pi_N(\mu)$ with respect to $d\alpha$ is also precisely the
average polynomial $\displaystyle \prod_{i=1}^N(\mu-x_i)$ with
respect to $d\Prob_{\alpha,N}$. Formula \eqref{0.5} appears
already in the work of Heine in the 1880's (see \cite{Szego}). Set
\begin{equation}\label{0.6}
  d\alphalm(t) \equiv
  \frac{\prod_{j=1}^\ell (\mu_j-t)}{\prod_{j=1}^m(\epsilon_j-t)}
  d\alpha(t), \qquad
  \ell,m\ge 0 ,
\end{equation}
($d\alpha^{[0,0]}(t)\equiv d\alpha(t)$), and let
$\pi_j^{[\ell,m]}(t)$ denote the $j$th monic orthogonal polynomial
with respect to $d\alphalm$. With this notation we see immediately
from \eqref{0.3}, \eqref{0.5} that $\bigl< \frac{\prod_{j=1}^K
D_N[\mu_j,H]}{\prod_{j=1}^M D_N[\epsilon_j,H]} \bigr>_\alpha$ is
proportional to $\pi^{[K-1,M]}_N(\mu_K)$ Using a classical
determinantal formula of Christoffel (see \cite{Szego}) for
$\pi^{[\ell,0]}_N(\mu)$ and a more recent formula of Uvarov
\cite{Uvarov} for $\pi^{[0,m]}_N(\mu)$, we are then led (see
Section \ref{Type1}. Formulae of Christoffel-Uvarov type) to
\eqref{BrezinHikamiFormula}, \eqref{averageOM} and
\eqref{AverageofRatios} in a rather straightforward way. Formula
\eqref{D} appears to have a different character from
\eqref{BrezinHikamiFormula}, \eqref{averageOM},
\eqref{AverageofRatios}, and relies on Lemma
\ref{LemmaChristoffel1} mentioned above, which computes the
integral of the product of two determinants: formula \eqref{E}
follows (see Section \ref{Type2}. Formulae of two-point function
type) by combining \eqref{D} with \eqref{BrezinHikamiFormula} and
\eqref{AverageofRatios}. In \cite{StrahovF2} the authors present a
variety of additional formulae for $\bigl< \frac{\prod_{j=1}^K
D_N[\mu_j,H]}{\prod_{j=1}^M D_N[\epsilon_j,H]} \bigr>_\alpha$ for
cases of $K$ and $M$ not covered by
\eqref{BrezinHikamiFormula}-\eqref{E}: we leave it to the
interested reader to verify that the method of this paper can also
be used to derive these formulae in a straightforward manner.

\begin{rems}
As is well-known (see e.g., \cite{Szego}), each measure $d\alpha$
gives rise to a tridiagonal operator
\begin{equation}\label{0.7}
  J=J(d\alpha) = \begin{pmatrix}
  a_1 & b_1 & 0 &\\
  b_1 & a_2 & b_2 &\\
  0   & b_2 & a_3 &\ddots \\
   & & \ddots & \ddots
  \end{pmatrix}, \qquad b_i>0
\end{equation}
with generalized eigenfunctions given by the orthonormal
polynomials
\begin{equation}\label{0.8}
  p_j(x)=c_j^{-1}\pi_j(x), \qquad j=0,1,\cdots,
\end{equation}
i.e.
\begin{equation}\label{0.9}
  b_{j-1} p_{j-1}(x)+a_jp_j(x)+b_jp_{j+1}(x)= xp_j(x),
  \qquad j\ge 1
\end{equation}
where $b_0\equiv 0$. Conversely, modulo certain essential
self-adjointness issues, $d\alpha$ is the spectral measure for $J$
in the cyclic subspace generated by $J$ and the vector
$e_1=(1,0,0,\cdots)^{T}$ (see, e.g., \cite{DeiftBook}). It follows
that the transformation of measures
\begin{equation}
  d\alpha \to d\alphalm
\end{equation}
leads to the transformation of operators
\begin{equation}
  J(d\alpha) \to J(d\alphalm).
\end{equation}
For appropriate choices of $\mu_1,\cdots,\mu_m$ and $\epsilon_1,
\cdots,\epsilon_\ell$, such transformations corresponding to
removing $m$ points from the spectrum of $J(d\alpha)$ and
inserting $\ell$ points: in the spectral theory literature, such
transformations are known as Darboux transformations. The formulae
in this paper clearly provide formulae for the generalized
eigenfunctions $p^{[\ell,m]}_j(x)$ of the Darboux-transformed
operator $J(d\alphalm)$, as well as the matrix entries,
$a_j^{[\ell,m]}$ and $b_j^{[\ell,m]}$, in terms of the
corresponding objects for $J(d\alpha)$. Again we leave the details
to the reader. Here the elementary formulae
\begin{equation}
  b_n^2(d\alpha) =
  \frac{n+1}{n+2}\frac{Z_n(d\alpha)Z_{n+2}(d\alpha)}{\bigl(Z_{n+1}(d\alpha)\bigr)^2},
  \qquad a_n(d\alpha)= \frac{d}{dt}\biggl|_{t=0} \log \frac{Z_n(d\alpha_t)}{Z_{n+1}(d\alpha_t)}
\end{equation}
where $d\alpha_t(x)=e^{tx}d\alpha(x)$, are useful.
\end{rems}

\begin{trems}\label{rem:techrem}
Formulae \eqref{BrezinHikamiFormula}-\eqref{E} clearly do not make
sense for all values of the parameters. In \emph{\textbf{all the
calculations that follow, we will assume that $d\alpha$ has
compact support, support($d\alpha$)$=[-Q,Q]$, say, and that the
$\mu_i$'s and $\epsilon_j$'s are distinct real numbers greater
than $Q$}}: under these assumptions, $d\alphalm(t)$ becomes, in
particular, a bona-fide measure, etc. By analytic continuation one
sees that the formulae remain true for complex values of
$\{\mu_i\}$ and $\{\epsilon_j\}$, as long as they remain distinct.
Furthermore, if the $\mu_i$'s and $\epsilon_j$'s are distinct, and
$Im(\epsilon_j)\neq 0$ for all $j$, then we can let $Q\to\infty$
and so the formulae are true for measures $d\alpha$ with unbounded
support. Finally we can, for example, let $\mu_j\to\mu_k$ for some
$j\neq k$, which leads to formulae involving derivatives of the
$\pi_j$'s, etc.
\end{trems}

\section{Formulae of Christoffel-Uvarov type}\label{Type1}

We use the notations $d\alpha$, $\pi_j$, $d\alphalm$, $\pilm_j$,
... of Section \ref{intro}. In addition, in all the calculations
that follow we assume that $d\alpha$, $\{\mu_j\}$,
$\{\epsilon_k\}$ satisfy the conditions described in Technical
Remark \ref{rem:techrem} above: the natural analytical
continuation of the formulae obtained to complex values of the
parameters, and the limit $Q\to\infty$, is left to the reader.

The following result of Christoffel (see \cite{Szego}) plays a
basic role in what follows.

\begin{lem}\label{LemmaChristoffel1}
Consider the measure
$d\alpha^{[\ell,0]}(t)=\prod\limits_{j=1}^\ell(\mu_j-t)\;d\alpha(t)$,
where $\ell=1,2,...$. Then the $\mbox{n}^{th}$ monic orthogonal
polynomial $\pi_{n}^{[\ell,0]}(t)$ associated with the new measure
$d\alpha^{[\ell,0]}(t)$ can be expressed as follows:
\begin{equation}\label{Christoffel1}
\pi_{n}^{[\ell,0]}(t)=\frac{1}{(t-\mu_1)\ldots (t-\mu_\ell)}\;
\frac{\left|\begin{array}{ccc}
  \pi_n(\mu_1) & \cdots & \pi_{n+\ell}(\mu_1) \\
  \vdots &  &  \\
  \pi_n(\mu_\ell) & \ldots & \pi_{n+\ell}(\mu_\ell) \\
  \pi_n(t) & \ldots & \pi_{n+\ell}(t)
\end{array}\right|}{\left|\begin{array}{ccc}
  \pi_{n}(\mu_1) & \ldots & \pi_{n+\ell-1}(\mu_1) \\
  \vdots &  &  \\
  \pi_{n}(\mu_\ell) & \ldots & \pi_{n+\ell-1}(\mu_\ell)
\end{array}\right|} .
\end{equation}
\end{lem}

\begin{proof}
Set
\begin{equation}
q_n^{[\ell,0]}(t)=  \left|\begin{array}{ccc}
  \pi_n(\mu_1) & \cdots & \pi_{n+\ell}(\mu_1) \\
  \vdots &  &  \\
  \pi_n(\mu_\ell) & \ldots & \pi_{n+\ell}(\mu_\ell) \\
  \pi_n(t) & \ldots & \pi_{n+\ell}(t)
\end{array}\right| .
\end{equation}
We note that $q_n^{[\ell,0]}(t)$ satisfies the condition $\int t^j
q_n^{[\ell,0]}(t)d\alpha(t)=0$ for all $j\in \{0,\ldots ,n-1\}$.
Also $q_n^{[\ell,0]}(\mu_j)=0$, $j=1,\cdots,\ell$, and so
$\frac{q_n^{[\ell,0]}(t)}{(\mu_1-t)\cdots(\mu_\ell-t)}$ is a
polynomial of degree at most $n$. Now observe that
\begin{equation}
\int t^j\left[\frac{q_n^{[\ell,0]}(t)}{(\mu_1-t)\ldots
(\mu_\ell-t)}\right]d\alpha^{[\ell,0]}(t)=0,\;\;0\leq j<n
\end{equation}
which means that $q_n^{[\ell,0]}(t)$ divided by the product
$(\mu_1-t)\ldots (\mu_\ell-t)$ is proportional to the $n^{th}$
monic orthogonal polynomial $\pi_n^{[\ell,0]}(t)$ associated with
the new measure $d\alpha^{[\ell,0]}(t)$. Now $q_n^{[\ell,0]}(t)$
cannot vanish for any $t=\mu_{\ell+1}>Q$, $\mu_{\ell+1}\notin
\{\mu_1,\cdots,\mu_\ell\}$. Indeed, if
$q_n^{[\ell,0]}(\mu_{\ell+1})=0$, then there exist
$\{\alpha_i\}_{i=0}^\ell$, not all zero, such that $p(t)\equiv
\sum_{i=0}^\ell \alpha_i \pi_{n+i}(t)$ vanishes at $\{
\mu_i\}_{i=1}^{\ell+1}$. Thus $\tilde{p}(t) \equiv p(t)/
\prod_{i=1}^{\ell+1} (\mu_i-t)$ is a polynomial of order $<n$, and
as above, $\tilde{p}(t)$ is orthogonal to $t^j$, $0\le j <n$, with
respect to the measure $d\alpha^{[\ell+1,0]}(t)$. Thus
$\tilde{p}(t)\equiv 0$ and hence $\alpha_0=\cdots=\alpha_\ell=0$,
which is a contradiction. Replacing $\ell$ by $\ell-1$, we
conclude that
\begin{equation}
  \left|\begin{array}{ccc}
  \pi_{n}(\mu_1) & \ldots & \pi_{n+\ell-1}(\mu_1) \\
  \vdots &  &  \\
  \pi_{n}(\mu_\ell) & \ldots & \pi_{n+\ell-1}(\mu_\ell)
\end{array}\right| \neq 0.
\end{equation}
Taking the limit $t\rightarrow \infty$ and noting that the
coefficient of the highest degree of $\pi_n^{[\ell,0]}(t)$ should
be equal to 1, we find the coefficient of proportionality and
establish formula (\ref{Christoffel1}).
\end{proof}

Representation (\ref{Christoffel1}) for the monic orthogonal
polynomials associated with the measure $d\alpha^{[\ell,0]}(t)$
immediately leads to the following result:

\begin{cor}\label{Product1}
The product of monic orthogonal polynomials
$\prod_{j=0}^\ell\pi_n^{[j,0]}(\mu_{j+1})$ defined with respect to
the different measures $d\alpha^{[j,0]}(t)\equiv (\mu_j-t)\cdots
(\mu_1-t)d\alpha(t)$ is given by the formula
\begin{equation}\label{EquationProduct1}
\prod_{j=0}^\ell\pi_n^{[j,0]}(\mu_{j+1})=\frac{1}{\triangle(\mu)}\;\left|
\begin{array}{ccc}
  \pi_n(\mu_1) & \cdots & \pi_{n+\ell}(\mu_1) \\
  \vdots &  &  \\
  \pi_n(\mu_{\ell+1}) & \cdots & \pi_{n+\ell}(\mu_{\ell+1})
\end{array}\right|
\end{equation}
where $\triangle(\mu)=\prod\limits_{\ell+1\geq i>j\geq
1}(\mu_i-\mu_j)$.
\end{cor}

We observe that Corollary (\ref{Product1}) gives the identity for
the average of products of random characteristic polynomials
obtained first by Brezin and Hikami \cite{BrezinH00}.

\begin{thm}\label{theorem1}
Let $D_N[\mu,H]$ be the characteristic polynomial of the Hermitian
matrix $H$. The following identity is valid:
\begin{equation}\label{BrezinHikamiFormula}
\left\langle\prod\limits_{j=1}^{L}D_N[\mu_j,H]\right\rangle_{\alpha}=
\frac{1}{\triangle(\mu)}\;\left|
\begin{array}{ccc}
  \pi_N(\mu_1) & \ldots & \pi_{N+L-1}(\mu_1) \\
  \vdots &  &  \\
  \pi_N(\mu_{L}) & \ldots & \pi_{N+L-1}(\mu_{L})
\end{array}\right|
\end{equation}
where the average is defined by \eqref{0.2}.
\end{thm}

\begin{proof}
To prove formula (\ref{BrezinHikamiFormula}) we use the
representation for the monic orthogonal polynomials in the case
$L=1$ given in \eqref{0.5},
\begin{equation}\label{IntegralMPolynomial}
\pi_{N}(\mu)=\frac{1}{Z_N}\int\ldots \int\prod\limits_{i=1}^N
(\mu-x_i)\triangle^2(x)d\alpha(x) .
\end{equation}
Let $Z_N^{[\ell,0]}$ be defined by
\begin{equation}
Z_N^{[\ell,0]}=\int\ldots\int\triangle^2(x)d\alpha^{[\ell,0]}(x),\;\;
\ell=1,2,\cdots.
\end{equation}
where
$d\alpha^{[\ell,0]}(x)=\prod\limits_{i=1}^{N}d\alpha^{[\ell,0]}(x_i)$.
With this notation, we have
\begin{equation}
\left\langle\prod\limits_{j=1}^{L}D_N[\mu_j,H]\right\rangle_{\alpha}=
\frac{Z_N^{[L,0]}}{Z_N}
=\frac{Z_N^{[L,0]}}{Z_N^{[L-1,0]}}\frac{Z_N^{[L-1,0]}}{Z_N^{[L-2,0]}}
\cdots\frac{Z_N^{[1,0]}}{Z_N}.
\end{equation}
Equation (\ref{IntegralMPolynomial}) implies that
$\pi_n^{[\ell-1,0]}(\mu_\ell)$ can be represented as the ratio
$Z_N^{[\ell,0]}/Z_N^{[\ell-1,0]}$, where $\pi_N^{[0,0]}(\mu)\equiv
\pi_N(\mu)$, and $Z_N^{[0,0]}\equiv Z_N$. Thus we obtain
\begin{equation}\label{AverageAsProduct1}
\left\langle\prod\limits_{j=1}^{L}D_N[\mu_j,H]\right\rangle_{\alpha}=
\prod\limits_{j=0}^{L-1}\pi_{N}^{[j,0]}(\mu_{j+1})
\end{equation}
The above equation together with Corollary (\ref{Product1}) proves
formula (\ref{BrezinHikamiFormula}).
\end{proof}

\begin{rems}
Notice (see equations (\ref{IntegralMPolynomial}) and
(\ref{AverageAsProduct1})) that the average of products of
characteristic polynomials can be rewritten as a product of
averages. Namely,
\begin{equation}
\left\langle\prod\limits_{j=1}^{L}D_N[\mu_j,H]\right\rangle_{\alpha}=
\prod\limits_{j=1}^L\left\langle
D_N[\mu_j,H]\right\rangle_{\alpha^{[j-1,0]}}
\end{equation}
where $\left\langle\ldots\right\rangle_{\alpha^{[j-1,0]}}$ means
the average defined by equation \eqref{0.2} but with respect to
the new measure $d\alpha^{[j-1,0]}(x)$, and $d\alpha(x)\equiv
d\alpha^{[0,0]}(x)$.
\end{rems}

\medskip

The formula of Christoffel (equation (\ref{Christoffel1})) enables
us to construct the orthogonal polynomials associated with the
measure
$d\alpha^{[\ell,0]}(t)=\prod\limits_{j=1}^\ell(\mu_j-t)d\alpha(t)$
in terms of the orthogonal polynomials associated with the measure
$d\alpha(t)$. Now we derive a formula due to Uvarov \cite{Uvarov}
expressing the monic orthogonal polynomials
 $\pi_n^{[0,m]}(t)$ associated with the measure
$d\alpha^{[0,m]}(t)=\prod\limits_{j=1}^{m}(\epsilon_j-t)^{-1}d\alpha(t)$,
again in terms of the monic orthogonal polynomials $\pi_n(t)$
associated with the measure $d\alpha(t)$.

\begin{lem}\label{lemmaChristoffel2}
Suppose $0\le m\le n$. The monic orthogonal polynomials
$\pi_n^{[0,m]}(t)$ associated with the measure
$d\alpha^{[0,m]}(t)$ can be expressed as ratios of determinants,
\begin{equation}\label{MonicOM}
\pi_{n}^{[0,m]}(t)= \frac{\left|\begin{array}{ccc}
  h_{n-m}(\epsilon_1) & \ldots & h_n(\epsilon_1) \\
  \vdots &  &  \\
  h_{n-m}(\epsilon_m) & \ldots & h_n(\epsilon_m) \\
  \pi_{n-m}(t) & \ldots & \pi_n(t)
\end{array}\right|}{\left|\begin{array}{ccc}
  h_{n-m}(\epsilon_1) & \ldots & h_{n-1}(\epsilon_1) \\
  \vdots &  &  \\
  h_{n-m}(\epsilon_m) & \ldots & h_{n-1}(\epsilon_m)
\end{array}\right|}.
\end{equation}
Here the $h_k(\epsilon_j)$'s are the Cauchy transformations of the
monic orthogonal polynomials $\pi_k(t)$,
\begin{equation}
h_k(\epsilon_j)=\frac{1}{2\pi i
}\int\frac{\pi_k(t)d\alpha(t)}{t-\epsilon_j}.
\end{equation}
\end{lem}

\begin{proof}
Set
\begin{equation}\label{1.14}
q_n^{[0,m]}(t)= \left|\begin{array}{ccc}
  h_{n-m}(\epsilon_1) & \ldots & h_n(\epsilon_1) \\
  \vdots &  &  \\
  h_{n-m}(\epsilon_m) & \ldots & h_n(\epsilon_m) \\
  \pi_{n-m}(t) & \ldots & \pi_n(t)
\end{array}\right|.
\end{equation}
Now $q_n^{[0,m]}(t)$ is proportional to the $n^{th}$ monic
orthogonal polynomial $\pi_n^{[0,m]}(t)$ with respect to the
measure $d\alpha^{[0,m]}(t)$. Indeed, first observe that
\begin{equation}\label{1.15}
  \int \frac{\qm_n(t)}{t-\epsilon_j} d\alpha(t) =0, \qquad j=1,\cdots,m.
\end{equation}
Also, for $0\le k<n$,
\begin{equation}\label{1/16}
  \frac{t^k}{\prod_{\ell=1}^m(\epsilon_\ell-t)} = \sum_{\ell=1}^m
  \frac{\beta_\ell}{\epsilon_\ell-t} + p(t)
\end{equation}
for suitable constants $\{\beta_\ell\}$ and for some polynomial of
degree $<n-m$. But for $0\le k<n$,
\begin{equation}\label{1.17}
  \int t^k \qm_n(t) d\alpham(t) = - \sum_{\ell=1}^m \beta_\ell
  \int \frac{\qm_n(t)}{t-\epsilon_\ell} d\alpha(t) + \int p(t)
  \qm_n(t) d\alpha(t).
\end{equation}
The terms in the sum are zero by \eqref{1.15} and the final
integral is zero by the construction \eqref{1.14} of $\qm_n(t)$
and the fact that deg $p(t)$ $<n-m$. Thus $\qm_n(t)$ is
proportional to $\pim_n(t)$. An argument similar to the proof in
Lemma \ref{LemmaChristoffel1} that
\begin{equation}
  \left|\begin{array}{ccc}
  \pi_{n}(\mu_1) & \ldots & \pi_{n+\ell-1}(\mu_1) \\
  \vdots &  &  \\
  \pi_{n}(\mu_\ell) & \ldots & \pi_{n+\ell-1}(\mu_\ell)
\end{array}\right| \neq 0,
\end{equation}
shows that the denominator in \eqref{MonicOM} does not vanish.
Letting $t\to\infty$ in \eqref{1.14}, and matching leading terms,
we prove Lemma \ref{lemmaChristoffel2}.
\end{proof}

\begin{rems}
In \cite{Uvarov}, Uvarov obtains formulae for $\pim_n(t)$ of type
\eqref{MonicOM} also in the case $m>n$. These formulae can be used
to obtain analogues of \eqref{averageOM} and
\eqref{AverageofRatios} below in the case $M>N$.
\end{rems}

\begin{rems}
As noted in \cite{StrahovF1, StrahovF2}, the Cauchy
transformations $h_k(\epsilon)$ of the $\pi_k$'s occur explicitly,
together with the $\pi_k$'s, in the solution of the
Fokas-Its-Kitaev Riemann-Hilbert problem for orthogonal
polynomials \cite{FIK}.
\end{rems}

Lemma (\ref{lemmaChristoffel2}) implies the following analogue of
the Christoffel formula for the Cauchy transforms of monic
orthogonal polynomials.

\begin{cor}
Let $h_k^{[0,m]}(\epsilon)$ be the Cauchy transform of the monic
polynomial $\pi_k^{[0,m]}(t)$ with respect to the measure
$d\alpha^{[0,m]}(t)$,
\begin{equation}\label{definitionofhkom}
h_k^{[0,m]}(\epsilon)=\frac{1}{2\pi i}
\int\frac{\pi_k^{[0,m]}(t)}{t-\epsilon}\;d\alpha^{[0,m]}(t) .
\end{equation}
Let also $0\leq m\leq n$. Then $h_n^{[0,m]}(\epsilon)$ has a
representation similar to that for the monic orthogonal
polynomials $\pi_n^{[l,0]}(t)$ (equation \eqref{Christoffel1}),
\begin{equation}\label{Christoffel2}
h_n^{[0,m]}(\epsilon)=\frac{(-1)^m}{(\epsilon-\epsilon_m)\ldots
(\epsilon-\epsilon_1)}\;\frac{\left|\begin{array}{ccc}
  h_{n-m}(\epsilon_1) & \ldots & h_n(\epsilon_1) \\
  \vdots &  &  \\
  h_{n-m}(\epsilon_m) & \ldots & h_n(\epsilon_m) \\
  h_{n-m}(\epsilon) & \ldots & h_n(\epsilon)
\end{array}\right|}{\left|\begin{array}{ccc}
  h_{n-m}(\epsilon_1) & \ldots & h_{n-1}(\epsilon_1) \\
  \vdots &  &  \\
  h_{n-m}(\epsilon_m) & \ldots & h_{n-1}(\epsilon_m)
\end{array}\right|} .
\end{equation}
\end{cor}

\begin{proof}
The above representation follows from formula (\ref{MonicOM}) and
from the fact that
\begin{equation}\label{TechnicalIdentity}
\frac{1}{(t-\epsilon_{m+1})\ldots
(t-\epsilon_1)}=\sum\limits_{j=1}^{m+1}\frac{1}{t-\epsilon_j}\prod\limits_{k\neq
j}\frac{1}{\epsilon_j-\epsilon_k} .
\end{equation}
Indeed we find from formula (\ref{MonicOM}) that
$h_n^{[0,m]}(\epsilon)$ is the ratio of the determinants. The
elements of the last row of the determinant in the numerator are
the integrals
\begin{equation}\frac{1}{2\pi
i}\;\int\frac{\pi_{n-k}(t)d\alpha(t)}{(t-\epsilon)(t-\epsilon_m)\ldots
(t-\epsilon_1)},\qquad 0\leq k\leq m\nonumber.
\end{equation}
Using identity (\ref{TechnicalIdentity}) and noting that the only
term
\begin{equation}
\frac{1}{t-\epsilon}\;\frac{1}{(\epsilon-\epsilon_m)\ldots
(\epsilon-\epsilon_1)} .
\end{equation}
of the sum (\ref{TechnicalIdentity}) contributes to the
determinant, (\ref{Christoffel2}) follows.
\end{proof}

Equation \eqref{Christoffel2} immediately implies the following
analogy of \eqref{EquationProduct1} for the $h^{[0,m]}_k$'s.

\begin{cor}
Let $0\le m\le n$. Then the product of the Cauchy transforms of
monic orthogonal polynomials with respect to the measures
$d\alpha^{[0,j]}(t)$, $0\leq j\leq m$ can be written as a
determinant,
\begin{equation}\label{CauchyProductAsDeterminant}
\prod\limits_{j=0}^{m}h_{n-m+j}^{[0,j]}(\epsilon_{j+1})=
\frac{(-1)^{\frac{m(m+1)}{2}}}{\triangle(\epsilon)}\; \left|
\begin{array}{ccc}
  h_{n-m}(\epsilon_1) & \ldots & h_n(\epsilon_1) \\
  \vdots &  &  \\
  h_{n-m}(\epsilon_{m+1}) & \ldots & h_{n}(\epsilon_{m+1})
\end{array}\right| .
\end{equation}
\end{cor}

Now we derive the identity for the average of the product of
inverse random characteristic polynomials.

\begin{thm}\label{theorem2}
Suppose $1\le M\le N$ and let $\gamma_n=-\frac{2\pi i}{c_n^2}$,
where $c_n$ is the normality constant defined by equation
\eqref{0.4}. Then we have the following formula
\begin{equation}\label{averageOM}
\left\langle\prod\limits_{j=1}^{M}D_N^{-1}\left[\epsilon_j,H\right]\right\rangle_{\alpha}
=(-1)^{\frac{M(M-1)}{2}}\;\frac{\prod_{j=N-M}^{N-1}\gamma_j}{\triangle(\epsilon)}
 \left|\begin{array}{ccc}
  h_{N-M}(\epsilon_1) & \ldots & h_{N-1}(\epsilon_1) \\
  \vdots &  &  \\
  h_{N-M}(\epsilon_{M}) & \ldots & h_{N-1}(\epsilon_{M})
\end{array}\right| .
\end{equation}
\end{thm}

\begin{proof}
When $M=1$, we use the identity \eqref{TechnicalIdentity} together
with \eqref{IntegralMPolynomial} and the relation (see, e.g.,
\cite{Szego})
\begin{equation}\label{1.23}
  \gamma_{n-1}= -2\pi in \frac{Z_{n-1}}{Z_n}
\end{equation}
to obtain
\begin{equation}\label{1.24}
  \bigl< D_N^{-1}[\epsilon, H]\bigr>_\alpha = \gamma_{N-1}
  h_{N-1}(\epsilon).
\end{equation}

We rewrite the average
 in  equation (\ref{averageOM}) as
follows:
\begin{equation}\label{1.25}
\left\langle\prod\limits_{j=1}^{M}D_N^{-1}\left[\epsilon_j,H\right]\right\rangle_{\alpha}
=\frac{Z_N^{[0,M]}}{Z_{N-1}^{[0,M-1]}}\;\frac{Z_{N-1}^{[0,M-1]}}{Z_{N-2}^{[0,M-2]}}\ldots
\frac{Z_{N-M}^{[0,0]}}{Z_{N}^{[0,0]}}
\end{equation}
where
\begin{equation}
Z_N^{[0,M]}=\int\ldots\int\triangle^2(x)d\alpha^{[0,M]}(x),
\end{equation}
$Z_N^{[0,0]}\equiv Z_N$ and $d\alpha^{[0,0]}(x)=d\alpha(x)$. The
following relation can be observed from equations \eqref{1.24} and
\eqref{1.23}:
\begin{equation}
\frac{Z_{N-K}^{[0,m]}}{Z_{N-K-1}^{[0,m-1]}}= -2\pi i
(N-K)\;h_{N-K-1}^{[0,m-1]}(\epsilon_{m}) .
\end{equation}
Inserting this relation in \eqref{1.25} we find
\begin{equation}
\left\langle\prod\limits_{j=1}^{M}D_N^{-1}\left[\epsilon_j,H\right]\right\rangle_{\alpha}
=\prod\limits_{j=1}^{M}\gamma_{N-j}h_{N-j}^{[0,M-j]}(\epsilon_{M-j+1}).
\end{equation}
Our result (\ref{averageOM}) immediately follows from the above
equation and formula (\ref{CauchyProductAsDeterminant}).
\end{proof}

We now repeat the above considerations for the case
\begin{equation}\label{rationalmeasure}
d\alphalm(t)=\frac{(\mu_1-t)\cdots
(\mu_\ell-t)}{(\epsilon_1-t)\cdots (\epsilon_m-t)}d\alpha(t) .
\end{equation}
The first result is a Christoffel type formula for the measure
(\ref{rationalmeasure}), which is due to Uvarov \cite{Uvarov}:

\begin{lem}\label{LemmaChristoffel3}
Suppose $0\le m\le n$. Then the monic orthogonal polynomials
$\pi_n^{[\ell,m]}(t)$'s with respect to the measure
$d\alpha^{\ell,m]}(t)$ have the following representation:
\begin{equation}\label{Christoffel3}
\pi_{n}^{[\ell,m]}(t)=\frac{1}{(t-\mu_\ell)\ldots (t-\mu_1)}\;
\frac{\left|\begin{array}{ccc}
  h_{n-m}(\epsilon_1) & \ldots & h_{n+\ell}(\epsilon_1) \\
  \vdots &  &  \\
  h_{n-m}(\epsilon_m) & \ldots & h_{n+\ell}(\epsilon_m)  \\
  \pi_{n-m}(\mu_1) & \ldots & \pi_{n+\ell}(\mu_1) \\
  \vdots &  &  \\
  \pi_{n-m}(\mu_\ell) & \ldots & \pi_{n+\ell}(\mu_\ell)\\
  \pi_{n-m}(t) & \ldots & \pi_{n+\ell}(t)
\end{array}\right|}{\left|\begin{array}{ccc}
  h_{n-m}(\epsilon_1) & \ldots & h_{n+\ell}(\epsilon_1) \\
  \vdots &  &  \\
  h_{n-m}(\epsilon_m) & \ldots & h_{n+\ell}(\epsilon_m)  \\
  \pi_{n-m}(\mu_1) & \ldots & \pi_{n+\ell}(\mu_1) \\
  \vdots &  &  \\
  \pi_{n-m}(\mu_\ell) & \ldots & \pi_{n+\ell}(\mu_\ell)
\end{array}\right|} .
\end{equation}
\end{lem}

\begin{proof}
As in the previous cases we define $\qlm_n(t)$ to be the
determinant in the numerator of \eqref{Christoffel3}. Observe that
\begin{equation}
q_n^{[\ell,m]}(\mu_1)=\ldots=q_n^{[\ell,m]}(\mu_\ell)=0
\end{equation}
and that
\begin{equation}
\int\frac{q_{n}^{[\ell,m]}(t)d\alpha(t)}{\epsilon_1-t}=\ldots
=\int\frac{q_{n}^{[\ell,m]}(t)d\alpha(t)}{\epsilon_m-t}=0 .
\end{equation}
The next steps are the same as in the proofs of Lemma
(\ref{LemmaChristoffel1}) and Lemma (\ref{lemmaChristoffel2}).
\end{proof}

\begin{cor}\label{OMAverageCorollary}
\begin{equation}\label{OMAverage}
\left\langle\prod\limits_{j=1}^KD_N[\mu_j,H]\right\rangle_{\alpha^{[0,M]}}=
\frac{1}{\triangle(\mu)}\;\frac{\left|\begin{array}{ccc}
  h_{N-M}(\epsilon_1) & \ldots & h_{N+K-1}(\epsilon_1) \\
  \vdots &  &  \\
  h_{N-M}(\epsilon_M) & \ldots & h_{N+K-1}(\epsilon_M)  \\
  \pi_{N-M}(\mu_1) & \ldots & \pi_{N+K-1}(\mu_1) \\
  \vdots &  &  \\
  \pi_{N-M}(\mu_K) & \ldots & \pi_{N+K-1}(\mu_K)
\end{array}\right|}{\left|\begin{array}{ccc}
  h_{N-M}(\epsilon_1) & \ldots & h_{N}(\epsilon_1) \\
  \vdots &  &  \\
  h_{N-M}(\epsilon_M) & \ldots &
  h_{N}(\epsilon_M)\end{array}\right|}.
\end{equation}
\end{cor}
\begin{proof}
Identity (\ref{OMAverage}) follows from equations
(\ref{AverageAsProduct1}) and (\ref{Christoffel3}) once we note
that equation (\ref{Christoffel3}) can be rewritten in a similar
manner as equation (\ref{EquationProduct1}).
\end{proof}
Finally we generalize Theorem (\ref{theorem1}) and Theorem
(\ref{theorem2}) and obtain a formula for the average of ratios of
characteristic polynomials.

\begin{thm}
Suppose $0\le M\le N$. Then the average of ratios of
characteristic polynomials of $N\times N$ Hermitian matrices $H$
is given by the following formula:
\begin{equation}\label{AverageofRatios}
  \biggl< \frac{\prod_{j=1}^KD_N[\mu_j,H]}{\prod_{j=1}^MD_N[\epsilon_j,H]}
\biggr>_{\alpha}=
\frac{(-1)^{\frac{M(M-1)}{2}}\prod\limits_{j=N-M}^{N-1}\gamma_j}{\triangle(\mu)\triangle(\epsilon)}
\left|\begin{array}{ccc}
  h_{N-M}(\epsilon_1) & \ldots & h_{N+K-1}(\epsilon_1) \\
  \vdots &  &  \\
  h_{N-M}(\epsilon_M) & \ldots & h_{N+K-1}(\epsilon_M)  \\
  \pi_{N-M}(\mu_1) & \ldots & \pi_{N+K-1}(\mu_1) \\
  \vdots &  &  \\
  \pi_{N-M}(\mu_K) & \ldots & \pi_{N+K-1}(\mu_K)
\end{array}\right| .
\end{equation}
\end{thm}

\begin{proof}
Let $\alpha^{[0,0]}\equiv\alpha$, $\Z^{[0,0]}_n\equiv Z_n$. Then
we have
\begin{equation}
\left\langle\frac{\prod_{j=1}^KD_N[\mu_j,H]}{\prod_{j=1}^MD_N[\epsilon_j,H]}
\right\rangle_{\alpha}=\frac{Z_N^{[K,M]}}{Z_N^{[0,0]}}=
\frac{Z_N^{[K,M]}}{Z_N^{[0,M]}}\frac{Z_N^{[0,M]}}{Z_N^{[0,0]}}
\end{equation}
i.e.
\begin{equation}
\left\langle\frac{\prod_{j=1}^KD_N[\mu_j,H]}{\prod_{j=1}^MD_N[\epsilon_j,H]}
\right\rangle_{\alpha}=
\left\langle\prod\limits_{j=1}^KD_N[\mu_j,H]
\right\rangle_{\alpha^{[0,M]}}\left\langle\prod\limits_{j=1}^MD_N^{-1}[\epsilon_j,H]
\right\rangle_{\alpha} .
\end{equation}
We use Corollary (\ref{OMAverageCorollary}) and Theorem
(\ref{theorem2}) to obtain formula (\ref{AverageofRatios}).
\end{proof}

\begin{rems}
Observe that formulae \eqref{BrezinHikamiFormula},
\eqref{averageOM} do not follow immediately as special cases of
\eqref{AverageofRatios}: some further algebraic manipulation is
required. Similarly, the process of adding and removing zeros is
clearly reciprocal. More precisely, given $\epsilon_1,\cdots,
\epsilon_\ell$, we can construct the polynomials
$\pi^{[0,\ell]}_n(t; d\alpha^{[0,\ell]})$ associated with the
measure $d\alpha^{[0,\ell]}(t)=
\bigl(\prod_{i=1}^{\ell}(\epsilon_i-t)^{-1}\bigr)dt$ by
\eqref{MonicOM}: We can then construct $\pi^{[\ell,0]}_n\bigl(t;
d(\alpha^{[0,\ell]})^{[\ell,0]}\bigr)$ with $\mu_i=\epsilon_i$,
inserting $\pi^{[0,\ell]}_n(t; d\alpha^{[0,\ell]})$ for $\pi_n(t)$
on the right-hand-side of \eqref{Christoffel1}. We should find
that
$\pi_n^{[\ell,0]}\bigl(t;d(\alpha^{[0,\ell]})^{[\ell,0]}\bigr) =
\pi_n(t;d\alpha)$. However, again, this relation is not
immediately clear, and requires further algebraic manipulation.
\end{rems}

\section{Formulae of two-point function type}\label{Type2}

The following integral version of the Binet-Cauchy formula is due
to Andr\'eief \cite{Andreief}, and plays a basic role in our
calculations.

\begin{lem}\label{lem2.1}
Let $(X,d\mu)$ be a measure space and suppose $f_i, g_j\in
L^2(X,d\mu)$ for $1\le i,j\le k$. Then
\begin{equation}\label{2.1}
\begin{split}
  & \int_X\cdots\int_X \det(f_i(x_j))_{1\le i,j\le k}
  \det(g_i(x_j))_{1\le i,j\le k} d\mu(x_1)\cdots d\mu(x_k) \\
  &\quad = k! \det \biggl( \int_X f_i(x) g_j(x) d\mu(x) \biggr)_{1\le i,j\le
  k}.
\end{split}
\end{equation}
\end{lem}

\begin{proof}
Set $c_{ij}= \int_X f_i(x) g_j(x) d\mu(x)$. Then
\begin{equation}\label{2.2}
\begin{split}
  & \int_X\cdots\int_X \det(f_i(x_j))_{1\le i,j\le k}
  \det(g_i(x_j))_{1\le i,j\le k} d\mu(x_1)\cdots d\mu(x_k) \\
  &= \sum_{\sigma, \tau \in S_k} \sgn(\sigma) \sgn(\tau)
  c_{\sigma(1)\tau(1)}\cdots c_{\sigma(k)\tau(k)} \\
  &= \sum_{\sigma} \sgn(\sigma) \sum_{\tau} \sgn(\tau\circ\sigma)
  c_{\sigma(1)\tau\circ\sigma(1)}\cdots c_{\sigma(k)\tau\circ\sigma(k)} \\
  &= \sum_{\sigma} (\sgn(\sigma))^2 \sum_{\tau} \sgn(\tau)
  c_{1\tau(1)}\cdots c_{k\tau(k)} \\
  &= k! \det(c_{ij})_{1\le i,j\le k}
\end{split}
\end{equation}
as desired. In \eqref{2.2} we used $\sgn(\tau\circ\sigma)
=(\sgn\tau)(\sgn\sigma)$ and the fact that
$c_{\sigma(1)\tau\circ\sigma(1)}\cdots
c_{\sigma(k)\tau\circ\sigma(k)} = c_{1\tau(1)}\cdots c_{k\tau(k)}$
for all $\sigma$.
\end{proof}

\begin{thm}\label{thm2.2}
Let $K\ge 1$. Then the following identity is valid:
\begin{equation}\label{D}
  \biggl< \prod_{j=1}^K D_N[\lambda_j,H] D_N[\mu_j,H]
  \biggr>_\alpha
  = \frac{C_{N,K}}{\Delta(\lambda)\Delta(\mu)} \det\bigl(
  W_{I,N+K}(\lambda_i, \mu_j) \bigr)_{1\le i,j\le K}
\end{equation}
where
\begin{equation}\label{2.4}
  W_{I,N+K}(x,y) = \frac{\piNK(x)\piNKone(y)- \piNK(y)\piNKone(y)}{x-y}
\end{equation}
and
\begin{equation}\label{2.4.5}
  C_{N,K}= \frac{\prod_{\ell=N}^{N+K-1} c_\ell^2}{(c_{N+K-1})^{2K}}
\end{equation}
where $c_\ell$ is again the norming constant for $\pi_\ell$ given
in \eqref{0.4}.
\end{thm}

\begin{proof}
Let $p_j(x)= c_j^{-1}\pi_j(x)$, $j\ge 0$, denote the orthonormal
polynomials with respect to $d\alpha$. From \eqref{0.2} we obtain
\begin{equation}\label{2.5}
  \biggl< \prod_{j=1}^K D_N[\lambda_j,H] D_N[\mu_j,H]
  \biggr>_\alpha
  = \frac1{Z_N \Delta(\lambda)\Delta(\mu)} \int\cdots\int
  \Delta(x,\lambda) \Delta(x,\mu) d\alpha(x).
\end{equation}
Adding columns, we see that the Vandermonde determinant
$\Delta(x,\lambda)$ has the form
\begin{equation}\label{2.5.5}
  \left|\begin{array}{cccc}
  \pi_{0}(x_1) & \pi_1(x_1) & \cdots & \pi_{N+K-1}(x_1) \\
  \vdots &  & & \\
  \pi_{0}(x_N) & \pi_1(x_N) & \cdots & \pi_{N+K-1}(x_N) \\
  \pi_0(\lambda_1) & \pi_1(\lambda_1) & \cdots &
  \pi_{N+K-1}(\lambda_1) \\
  \vdots & & & \\
  \pi_0(\lambda_K) & \pi_1(\lambda_K) & \cdots &
  \pi_{N+K-1}(\lambda_K)
\end{array}\right|
\end{equation}
and similarly for $\Delta(x,\mu)$. Here
$\pi_j(t)=\pi^{[0,0]}_j(t)$. The determinant $\Delta(x,\lambda)$
can be evaluated by a Lagrange expansion of the form
\begin{equation}\label{2.6}
  \sum_{0\le i_1<i_2<\cdots<i_k\le N+K-1} \sigma_{i_1,\cdots,i_K}
  \left|\begin{array}{ccc}
  \pi_{i_1}(\lambda_1) & \cdots & \pi_{i_K}(\lambda_1) \\
  \vdots &  & \\
  \pi_{i_1}(\lambda_K) & \cdots & \pi_{i_K}(\lambda_K)
\end{array}\right|
  \left|\begin{array}{ccc}
  \pi_{j_1}(x_1) & \cdots & \pi_{j_N}(x_1) \\
  \vdots &  & \\
  \pi_{j_1}(x_N) & \cdots & \pi_{j_N}(x_N)
\end{array}\right|
\end{equation}
where $\sigma_{i_1,\cdots,i_K}=\pm 1$ is an appropriate signature
and $\{ (j_1,\cdots,j_N) : 0\le j_1<j_2<\cdots<j_N\le N+K-1 \}$ is
the complement of $\{ i_1,\cdots, i_K\}$ in $\{0,1,\cdots,
N+K-1\}$. Multiplying \eqref{2.6} by a similar expansion for
$\Delta(x,\mu)$, and inserting in \eqref{2.5}, we obtain a sum of
terms of the form
\begin{equation}\label{2.6.1}
  \int\cdots\int  \left|\begin{array}{ccc}
  \pi_{j_1}(x_1) & \cdots & \pi_{j_K}(x_1) \\
  \vdots &  & \\
  \pi_{j_1}(x_N) & \cdots & \pi_{j_K}(x_N)
\end{array}\right|
 \left|\begin{array}{ccc}
  \pi_{j_1'}(x_1) & \cdots & \pi_{j_N'}(x_1) \\
  \vdots &  & \\
  \pi_{j_1'}(x_N) & \cdots & \pi_{j_N'}(x_N)
\end{array}\right| d\alpha(x)
\end{equation}
which is equal by Lemma \ref{lem2.1} to $N! \det\bigl( \int
\pi_{j_i'}(x)\pi_{j_k}(x)d\alpha(x) \bigr)_{1\le i,k\le N} = N!
\det(\delta_{j_i'j_k} c_{j_k}^2)_{1\le i,k\le N}$. From this we
see that
\begin{equation}\label{2.6.2}
\begin{split}
  & \biggl< \prod_{j=1}^K D_N[\lambda_j,H] D_N[\mu_j,H]
  \biggr>_\alpha \\
  & = \frac{N!}{Z_N\Delta(\lambda)\Delta(\mu)}
  \sum_{0\le i_1<\cdots<i_k\le N+K-1} \sigma_{i_1,\cdots,i_K}^2
  \left|\begin{array}{ccc}
  \pi_{i_1}(\lambda_1) & \cdots & \pi_{i_K}(\lambda_1) \\
  \vdots &  & \\
  \pi_{i_1}(\lambda_K) & \cdots & \pi_{i_K}(\lambda_K)
\end{array}\right| \\
  &\qquad \times
  \prod_{k=1}^N c_{j_k}^2
  \left|\begin{array}{ccc}
  \pi_{i_1}(\mu_1) & \cdots & \pi_{i_K}(\mu_1) \\
  \vdots &  & \\
  \pi_{i_1}(\mu_K) & \cdots & \pi_{i_K}(\mu_K)
\end{array}\right| \\
  &= \frac{N! \prod_{q=N}^{N+K-1}
  c_q^2}{Z_N\Delta(x,\lambda)\Delta(x,\mu)}
  \sum_{0\le i_1<\cdots<i_k\le N+K-1} \det\bigl( p_{i_j}(\lambda_k)
  \bigr)_{1\le j,k\le K} \det\bigl( p_{i_j}(\mu_k)
  \bigr)_{1\le j,k\le K} \\
  &= \frac{N! \prod_{q=N}^{N+K-1}
  c_q^2}{Z_N\Delta(x,\lambda)\Delta(x,\mu)}
  \det\biggl( \sum_{0\le i\le N+K-1} p_{i}(\lambda_j) p_{i}(\mu_k)
  \biggr)_{1\le j,k\le K}
\end{split}
\end{equation}
where the last line follows by applying Lemma \ref{lem2.1} to the
discrete measure $d\mu=\sum_{i=0}^{N+K-1} \delta_i$. But by the
Christoffel-Darboux formula
\begin{equation}\label{2.6.3}
  \sum_{0\le i\le N+K-1} p_{i}(\lambda_j) p_{i}(\mu_k)
  = \frac{\piNK(\lambda_j)\piNKone(\mu_k)- \piNK(\mu_k)\piNKone(\lambda_j)}{\lambda_j-\mu_k}
\end{equation}
which then implies \eqref{D} as $Z_N=N!
\prod_{\ell=0}^{N-1}c_\ell^2$ (see, e.g. \cite{Szego}).
\end{proof}

\begin{thm}\label{thm2.3}
Suppose $1\le K\le N$. Then the following identity is valid:
\begin{equation}\label{E}
  \biggl< \prod_{j=1}^K \frac{D_N[\mu_i,H]}{D_N[\epsilon_j,H]}
  \biggr>_\alpha
  = (-1)^{K(K-1)/2} \gamma_{N-1}^K \frac{\Delta(\epsilon,\mu)}{\Delta^2(\epsilon)\Delta^2(\mu)} \det\bigl(
  W_{II,N}(\epsilon_i, \mu_j) \bigr)_{1\le i,j\le K}
\end{equation}
where
\begin{equation}\label{2.7.1}
  W_{II,N}(x,y) = \frac{h_N(\epsilon)\pi_{N-1}(\mu)- h_{N-1}(\epsilon)\pi_N(\mu)}{\epsilon-\mu}
\end{equation}
and again $h_k(\epsilon) = \frac1{2\pi i} \int
\frac{\pi_k(t)d\alpha(t)}{t-\epsilon}$ is the Cauchy transform of
$\pi_k(t)$ and $\gamma_{N-1}=-2\pi i/C_{N-1}^2$.
\end{thm}

Observe first that by linearity
\begin{equation}\label{2.7.2}
\begin{split}
   \left|\begin{array}{ccc}
  h_{N-M}(\epsilon_1) & \cdots & h_{N+L-1}(\epsilon_1) \\
  \vdots & & \\
  h_{N-M}(\epsilon_M) & \cdots & h_{N+L-1}(\epsilon_M) \\
  \pi_{N-M}(\mu_1) & \cdots & \pi_{N+L-1}(\mu_1) \\
  \vdots &  & \\
  \pi_{N-M}(\mu_1) & \cdots & \pi_{N+L-1}(\mu_L)
\end{array}\right|
  = & \int\cdots\int \frac{d\alpha(\lambda)}{(2\pi i)^M
  \prod_{j=1}^M(\lambda_j-\epsilon_j)} \\
  & \times
  \left|\begin{array}{ccc}
  \pi_{N-M}(\lambda_1) & \cdots & \pi_{N+L-1}(\lambda_1) \\
  \vdots & & \\
  \pi_{N-M}(\lambda_M) & \cdots & \pi_{N+L-1}(\lambda_M) \\
  \pi_{N-M}(\mu_1) & \cdots & \pi_{N+L-1}(\mu_1) \\
  \vdots &  & \\
  \pi_{N-M}(\mu_1) & \cdots & \pi_{N+L-1}(\mu_L)
\end{array}\right| .
\end{split}
\end{equation}
Inserting \eqref{AverageofRatios} on the left-hand-side, and using
\eqref{EquationProduct1} to re-express the integrand on the
right-hand-side, we obtain the following result, which is of
independent interest. The result expresses averages of ratios of
characteristic polynomials in terms of averages of products of
such polynomials.

\begin{prop}
Let $1\le M\le N$. Then
\begin{equation}\label{2.8}
\begin{split}
    & \biggl< \frac{\prod_{j=1}^L D_N[\mu_i,H]}{\prod_{j=1}^M D_N[\epsilon_j,H]}
  \biggr>_\alpha
  = \frac{(-1)^{M(M-1)/2}\prod_{j=N-M}^{N-1}
  \gamma_j}{\Delta(\mu)\Delta(\epsilon)} \\
  &\quad \times
  \int\cdots\int \frac{d\alpha(\lambda)}{(2\pi i)^M
  \prod_{j=1}^M(\lambda_j-\epsilon_j)}
  \Delta(\lambda, \mu)
  \biggl< \prod_{j=1}^M D_{N-M}[\lambda_j,H] \prod_{j=1}^L D_{N-M}[\mu_j,H]
  \biggr>_\alpha .
\end{split}
\end{equation}
\end{prop}

\begin{proof}[Proof of Theorem \ref{thm2.2}]
For $M=L=K\le N$, by \eqref{2.8} and \eqref{D},
\begin{equation}\label{2.8.1}
\begin{split}
  & \frac{\Delta(\mu)\Delta(\epsilon)}{(-1)^{K(K-1)/2}\prod_{j=N-K}^{N-1}
  \gamma_j}
  \biggl< \frac{\prod_{j=1}^K D_N[\mu_i,H]}{\prod_{j=1}^K D_N[\epsilon_j,H]}
  \biggr>_\alpha \\
  &=
  \int\cdots\int \frac{d\alpha(\lambda)}{(2\pi i)^M
  \prod_{j=1}^M(\lambda_j-\epsilon_j)}
  C_{N-K,K} \prod_{i=1}^K\prod_{j=1}^K(\mu_i-\lambda_j)
  \det\bigl( W_{I,N}(\lambda_i,\mu_j) \bigr)_{1\le i,j\le K}.
\end{split}
\end{equation}
But
\begin{equation}\label{2.8.2}
\begin{split}
  &\frac1{2\pi i} \int
  \frac{d\alpha(\lambda_j)}{\lambda_j-\epsilon_j}
  \prod_{i=1}^K(\mu_i-\lambda_j)
  \frac{\pi_N(\lambda_j)\pi_{N-1}(\mu_k)-\pi_{N-1}(\lambda_j)\pi_N(\mu_k)}{\lambda_j-\mu_k}\\
  & = \frac1{2\pi i} \int d\alpha(\lambda_j) \biggl( 1- \frac{\mu_1-\epsilon_j}{\lambda_j-\epsilon_j}
  \biggr) \biggl( \prod_{\substack{i=2 \\ i\neq k}}^K
  (\mu_i-\lambda_j)\biggr)
  \bigl(
  \pi_N(\lambda_j)\pi_{N-1}(\mu_k)-\pi_{N-1}(\lambda_j)\pi_N(\mu_k)\bigr)
  \\
  &= -\frac1{2\pi i} \int d\alpha(\lambda_j)
  \frac{\mu_1-\epsilon_j}{\lambda_j-\epsilon_j} \biggl( \prod_{\substack{i=2 \\ i\neq k}}^K
  (\mu_i-\lambda_j)\biggr) \bigl(
  \pi_N(\lambda_j)\pi_{N-1}(\mu_k)-\pi_{N-1}(\lambda_j)\pi_N(\mu_k)\bigr)
\end{split}
\end{equation}
as $\int d\alpha(\lambda_j) \lambda^\ell_j \pi_{N-1}(\lambda_j) =
\int d\alpha(\lambda_j) \lambda^\ell_j \pi_{N}(\lambda_j) = 0$ for
$0\le \ell\le K-2< N-1$. Continuing in this way, the integral
reduces to
$\prod_{i=1}^K(\mu_i-\epsilon_j)W_{II,N}(\epsilon_i,\mu_k)$. Thus
we find
\begin{equation}\label{2.8.3}
\begin{split}
  \frac{\Delta(\mu)\Delta(\epsilon)}{(-1)^{K(K-1)/2}\prod_{j=N-K}^{N-1}
  \gamma_j}
  \biggl< \frac{\prod_{j=1}^K D_N[\mu_i,H]}{\prod_{j=1}^K D_N[\epsilon_j,H]}
  \biggr>_\alpha
  =
  \frac{\Delta(\epsilon,\mu)}{\Delta(\epsilon)\Delta(\mu)}
  \det\bigl(
  W_{I,N+K}(\lambda_i, \mu_k) \bigr)_{1\le i,k\le K}
\end{split}
\end{equation}
and \eqref{E} follows.
\end{proof}

\medskip
\noindent {\bf Acknowledgments.} The authors would like to thank
Jeff Geronimo for useful conversations and for pointing out the
paper of U. B. Uvarov. The authors would also like to thank Nick
Witte for many useful remarks. The work of the first author was
supported in part by NSF Grant \# DMS-0208577. The work of the
second author was supported in part by NSF Grant \# DMS-0296084
and by the Institute for Advanced Study in Princeton. The work of
the third author was supported in part by EPSRC Grant GR/13838/01
``Random Matrices close to Unitary or Hermitian".


\begin{thebibliography}{10}

\bibitem{AndreevS}
A.V.~Andreev and V.D.~Simons.
\newblock Correlators of spectral determinants in quantum chaos.
\newblock {\em Phys. Rev. Lett.}, 75:2304--2307, 1995.

\bibitem{Andreief}
C.~Andr\'eief.
\newblock Note sur une relation les int{\'e}grales d{\'e}finies des produits
  des fonctions.
\newblock {\em M{\'e}m. de la Soc. Sci. Bordeaux}, 2:1-14, 1883.

\bibitem{BrezinH00}
E.~Brezin and S.~Hikami.
\newblock Characteristic polynomials of random matrices.
\newblock {\em Comm. Math. Phys.}, 214:111--135, 2000.

\bibitem{DeiftBook}
P.~Deift.
\newblock {\em Orthogonal polynomials and random matrices: a
  {R}iemann-{H}ilbert approach}, volume~3 of {\em Courant {L}ecture {N}otes in
  {M}athematics}.
\newblock AMS, Providence, 2000.



\bibitem{DKMVZ3}
P.~Deift, T.~Kriecherbauer, K.~McLaughlin, S.~Venakides, and
X.~Zhou.
\newblock Uniform asymptotics for polynomials orthogonal with respect to
  varying exponential weights and applications to universality questions in
  random matrix theory.
\newblock {\em Comm. Pure Appl. Math.}, 52(11):1335--1425, 1999.

\bibitem{DKMVZ2}
P.~Deift, T.~Kriecherbauer, K.~McLaughlin, S.~Venakides, and
X.~Zhou.
\newblock Strong asymptotics of orthogonal polynomials with respect to
  exponential weights.
\newblock {\em Comm. Pure Appl. Math.}, 52(12):1491--1552, 1999.

\bibitem{DVZ}
P.~Deift, S.~Venakides, and X.~Zhou.
\newblock New results in small dispersion {K}d{V} by an extension of the
  steepest descent method for {R}iemann-{H}ilbert problems.
\newblock {\em Internat. Math. Res. Notices}, 6:285--299, 1997.

\bibitem{DZ1}
P.~Deift and X.~Zhou.
\newblock A steepest descent method for oscillatory {R}iemman-{H}ilbert
  problems. Asymptotics for the {MKdV} equation.
\newblock {\em Ann. of Math.}, 137:295--368, 1993.

\bibitem{FIK}
A.~Fokas, A.~Its, and V.~Kitaev.
\newblock Discrete {P}ainlev{\'e} equations and their appearance in quantum
  gravity.
\newblock {\em Comm. Math. Phys.}, 142:313--344, 1991.

\bibitem{ForresterW}
P.~Forrester and N.~Witte.
\newblock Applications of the tau-function theory of {P}ainlev{\'e} equations
  to random matrices: {PIV}, {PII} and the {GUE}.
\newblock {\em Comm. Math. Phys.}, 219(2):357--398, 2001.

\bibitem{StrahovF2}
E.~Strahov and Y.~Fyodorov.
\newblock Universal results for correlations of characteristic polynomials:
  {R}iemann-{H}ilbert approach.
\newblock \textbf{arXiv:math-ph/0210010}, 2002.

\bibitem{StrahovF1}
Y.~Fyodorov and E.~Strahov.
\newblock An exact formula for general spectral correlation functions of random
  matrices.
\newblock {\em J. Phys. A: Math. Gen.}, 36, 2003.

\bibitem{KeatingHPO01}
C.~Hughes, J.~Keating, N.~O'Connell.
\newblock On the characteristic polynomials of a random unitary matrix.
\newblock {\em Comm. Math. Phys.}, 220(2):429--451, 2001.

\bibitem{KeatingSH00}
C.~Hughes, J.~Keating, and N.~O'Connell.
\newblock Random matrix thoery and the derivative of the {R}iemann-zeta
  function.
\newblock {\em R. Soc. Lond. Proc. Ser. A Math. Phys. Eng. Sci.}, 456:2611--2627, 2000.

\bibitem{KeatingS00}
J.~Keating and N.~Snaith.
\newblock Random matrix theory and $\zeta(\frac12+it)$.
\newblock {\em Comm. Math. Phys.}, 214:57--89, 2000.

\bibitem{Mehta91}
M.~Mehta.
\newblock {\em Random matrices}.
\newblock Academic Press, San Diego, Second Edition, 1991.

\bibitem{MehtaN01}
M.~Mehta and J-M. Normand.
\newblock Moments of the characteristic polynomial in the three ensembles of
  random matrices.
\newblock {\em J. Phys. A: Math. Gen.}, 34:4627--4639, 2001.

\bibitem{Szego}
G.~Szeg{\"o}.
\newblock {\em Orthogonal Polynomials}, volume~23 of {\em American Mathematical
  Society, Colloquium Publications}.
\newblock AMS, Providence, R.I., Fourth Edition, 1975.

\bibitem{Uvarov}
V.~B. Uvarov.
\newblock The connection between systems of polynomials orthogonal with respect
  to different distribution functions.
\newblock {\em USSR Comput. Math. and Math. Phys.}, 9 (part 2):25-36, 1969.

\end{thebibliography}

\end{document}